\documentclass{aa}
\usepackage{psfig}

\def\ltsima{$\; \buildrel < \over \sim \;$}
\def\lsim{\lower.5ex\hbox{\ltsima}}
\def\gtsima{$\; \buildrel > \over \sim \;$}
\def\gsim{\lower.5ex\hbox{\gtsima}}
\def\fdeg{\hbox{$\,.\!\!^{\circ}$}}

\begin{document}

\title{Physics of the GRB 030328 afterglow and its environment\thanks{Based 
on observations collected with: the Nordic Optical Telescope (NOT)
operating on the island of La Palma in the Spanish Observatorio del Roque
de los Muchachos of the Instituto de Astrof\'{i}sica de Canarias (Spain),
the 1.54 Danish telescope and the 2.2m ESO/MPG telescope (the latter under 
ESO Program ID: 70.D-0523) both operating at ESO-La Silla (Chile), the 
VLT plus FORS1 (ESO Program IDs: 70.D-0523 and 70.D-0531) at 
ESO-Paranal (Chile), the 40-inch of Siding Spring Observatory (Australia), 
the 1m Sampurnanand telescope at the ARIES observatory (India),
and the 1m NOFS telescope of USNO in Flagstaff (USA)}
}

\titlerunning{Physics of the GRB 030328 afterglow and its environment}
\authorrunning{E. Maiorano et al.}

\author{E. Maiorano\inst{1,2,3}
\and
N. Masetti\inst{1}
\and
E. Palazzi\inst{1}
\and
S. Savaglio\inst{4}
\and
E. Rol\inst{5}
\and
P.M. Vreeswijk\inst{6}
\and
E. Pian\inst{1,2}
\and
P.A. Price\inst{7,8}
\and
B.A. Peterson\inst{8}
\and
M. Jel\'{i}nek\inst{9}
\and
L. Amati\inst{1}
\and
M.I. Andersen\inst{10}
\and
A.J. Castro-Tirado\inst{9}
\and
J.M. Castro Cer\'on\inst{11}
\and
A. de Ugarte Postigo\inst{9}
\and
F. Frontera\inst{1,12}
\and
A.S. Fruchter\inst{13}
\and
J.P.U. Fynbo\inst{11}
\and
J. Gorosabel\inst{9}
\and
A.A. Henden\inst{14}
\and
J. Hjorth\inst{11}
\and
B.L. Jensen\inst{11}
\and
S. Klose\inst{15}
\and
C. Kouveliotou\inst{16}
\and
G. Masi\inst{17}
\and
P. M\o ller\inst{18}
\and
L. Nicastro\inst{1}
\and
E.O. Ofek\inst{19}
\and
S.B. Pandey\inst{9,20}
\and
J. Rhoads\inst{13}
\and
N.R. Tanvir\inst{21}
\and
R.A.M.J. Wijers\inst{22}
\and
E.P.J. van den Heuvel\inst{22}
}

\institute{
INAF -- Istituto di Astrofisica Spaziale e Fisica Cosmica di Bologna, 
via Gobetti 101, I-40129 Bologna, Italy
\and
INAF -- Osservatorio Astronomico di Trieste, via G.B. Tiepolo 11, I-34131
Trieste, Italy
\and
Dipartimento di Astronomia, Universit\`a di Bologna, via Ranzani 1,
I-40126 Bologna, Italy
\and
The Johns Hopkins University, 3400 North Charles Street, Baltimore, MD
21218, USA
\and
Department of Physics and Astronomy, University of Leicester, University
Road, Leicester, LE1 7RH, United Kingdom
\and
European Southern Observatory, Casilla 19001, Santiago 19, Chile
\and
Institute for Astronomy, University of Hawaii, 2680 Woodlawn Drive,
Honolulu, HI 96822, USA
\and
Research School of Astronomy and Astrophysics, Australian National 
University, via Cotter Road, Weston, ACT 2611, Australia
\and
Instituto de Astrof\'{\i}sica de Andaluc\'{\i}a (IAA-CSIC),
P.O. Box 03004, E-18080 Granada, Spain
\and
Astrophysikalisches Institut, D-14482 Potsdam, Germany
\and
Dark Cosmology Centre, Niels Bohr Institute, University of Copenhagen, 
Juliane Maries Vej 30, DK--2100 Copenhagen \O, Denmark
\and
Dipartimento di Fisica, Universit\`a di Ferrara, via Saragat 1, I-44100
Ferrara, Italy
\and
Space Telescope Science Institute, 3700 San Martin Drive, Baltimore, MD
21218, USA
\and
Universities Space Research Association / U.S. Naval Observatory, P.O.
Box 1149, Flagstaff, AZ 86002, USA
\and
Th\"uringer Landessternwarte Tautenburg, D-07778 Tautenburg, Germany
\and
NASA MSFC, SD-50, Huntsville, AL 35812, USA
\and
Dipartimento di Fisica, Universit\`a di Roma ``Tor Vergata", 
via della Ricerca Scientifica 1, I-00133 Rome, Italy
\and
European Southern Observatory, Karl Schwarzschild-Strasse 2, D-85748
Garching, Germany
\and
School of Physics and Astronomy and the Wise Observatory, University of 
Tel-Aviv, Tel-Aviv 69978, Israel
\and
ARIES Observatory, Manora Peak, Naini Tal -- 263129 Uttaranchal, India
\and
Department of Physical Sciences, University of Hertfordshire,
College Lane, Hatfield, Herts AL10 9AB, UK
\and
Institute of Astronomy ``Anton Pannekoek", University of Amsterdam,
Kruislaan 403, NL-1098 SJ Amsterdam, The Netherlands
}

\offprints{E. Maiorano, {\tt maiorano@iasfbo.inaf.it}}
\date{Received 20 December 2005; Accepted 19 March 2006}

\abstract{We report on the photometric, spectroscopic and polarimetric 
monitoring of the optical afterglow of Gamma--Ray Burst (GRB) 030328 
detected by {\it HETE-2}. Photometry, collected at 7 different telescopes, 
shows that a smoothly broken powerlaw decay, with indices $\alpha_1 = 0.76 
\pm 0.03$, $\alpha_2 = 1.50 \pm 0.07$ and a break at $t_b$ = 0.48 $\pm$ 
0.03 days after the GRB, provides the best fit of the optical afterglow 
decline. This shape is interpreted as due to collimated emission, for 
which we determine a jet opening angle $\theta_{\rm jet} \sim$ 3$\fdg$2. 
An achromatic bump starting around $\sim$0.2 d after the GRB is possibly 
marginally detected in the optical light curves. Optical spectroscopy 
shows the presence of two rest-frame ultraviolet metal absorption systems 
at $z$ = 1.5216 $\pm$ 0.0006 and at $z$ = 1.295 $\pm$ 0.001, the former 
likely associated with the GRB host galaxy. Analysis of the absorption 
lines at $z$ = 1.5216 suggests that the host of this GRB may be a Damped 
Lyman-$\alpha$ Absorber. The optical $V$-band afterglow appears polarized, 
with $P$ = (2.4 $\pm$ 0.6) \% and $\theta$ = $170^{\circ} \pm 7^{\circ}$, 
suggesting an asymmetric blastwave expansion. An X--ray-to-optical 
spectral flux distribution of the GRB 030328 afterglow was obtained at 
0.78 days after the GRB and fitted using a broken powerlaw, with an 
optical spectral slope $\beta_{\rm opt} = 0.47 \pm 0.15$, and an X--ray 
slope $\beta_{\rm X} = 1.0 \pm 0.2$. The discussion of these results in 
the context of the ``fireball model'' shows that the preferred scenario 
for this afterglow is collimated structured jet with fixed opening angle 
in a homogeneous medium.
 
}

\maketitle


\section{Introduction}

Long Gamma-Ray Bursts (GRBs) are rapid and powerful high-energy events 
lasting more than $\sim$2 s and up to $\sim$1000 s (e.g., Kouveliotou et 
al. 1993). The detection of their longer-wavelength counterparts (Costa et 
al. 1997; van Paradijs et al. 1997; Guarnieri et al. 1997; Frail et al. 
1997), has made it clear that multiwavelength studies of these phenomena 
are critical in order to constrain the parameters of the emission models 
and of the circumburst medium (see e.g. Galama et al. 1998a; Palazzi et 
al. 1998; Masetti et al. 1999).

The discovery that the optical emission of GRB afterglows is polarized at 
the level of a few percent (Covino et al. 1999; Wijers et al. 1999) points 
to the relevance of asymmetries in the afterglow emission (Rol et al. 
2000, 2003a; Greiner et al. 2003a) possibly induced by the presence of a 
collimated jet (Sari 1999; Ghisellini \& Lazzati 1999). However, there 
have been few polarimetric measurements of GRB optical afterglows so far, 
due to the rapid fading of these sources and to the relatively low level 
of the polarized signal, requiring timely observations with large 
telescopes.

Due to their high intrinsic brightness in the early emission phases, long 
GRBs are often used as a probe with which to sample the metal content of 
their host galaxies (e.g., Savaglio et al. 2003) and thus are a powerful 
tool with which to map the metallicity evolution in the Universe at very 
high ($z >$ 1) redshifts. This is an important parameter with which to 
characterize the metallicity content of the environments where the massive 
progenitors of GRBs originate (Galama et al. 1998b; Iwamoto et al. 1998; 
Hjorth et al. 2003; Stanek et al. 2003; Matheson et al. 2003; Malesani et 
al. 2004; Woosley \& Heger 2006).


\begin{figure}[th!]
\psfig{file=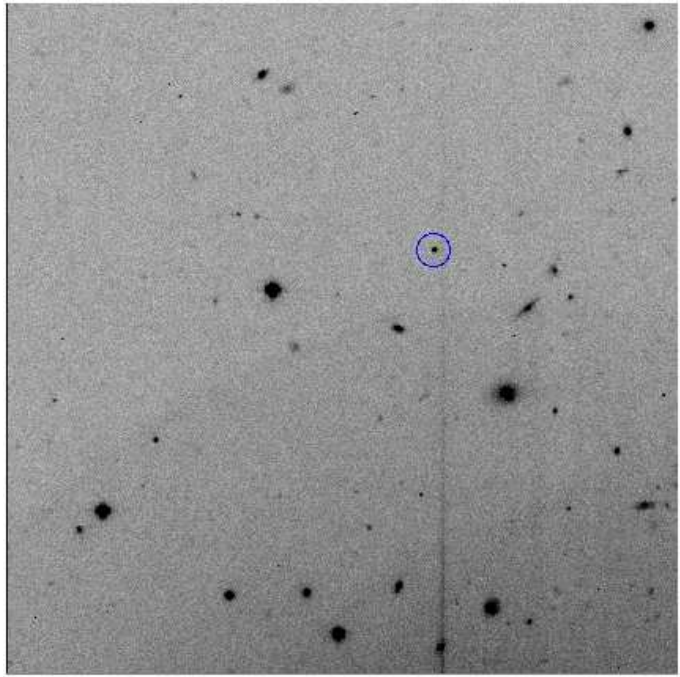,width=9cm}
\caption{$R$-band image of the field of the GRB 030328 acquired at ESO-La 
Silla with the 2.2m ESO/MPG plus WFI on 2003 Mar 29.273 UT. North is at 
top, East is to the left. The field size is about $2' \times 2'$. The OT 
is clearly detected and indicated by the circle.}
\end{figure}


GRB 030328 was a long (T$_{90}$ $\sim$ 100 s at 30--400 keV), bright 
(fluence: $S_\gamma$ = $3.0 \times 10^{-5}$ erg cm$^{-2}$; peak flux: 
$F_\gamma$ = $7.3 \times 10^{-7}$ ergs cm$^{-2}$ s$^{-1}$) GRB detected on 
2003 Mar 28.4729 UT, by the FREGATE, WXM, and SXC instruments onboard {\it 
HETE-2}, and rapidly localized with sub-arcminute accuracy (Villasenor et 
al. 2003). About $\sim$1 hour after the GRB, its optical afterglow was 
detected by the 40-inch Siding Spring Observatory (SSO) telescope in 
Australia at coordinates (J2000) RA = 12$^{\rm h}$ 10$^{\rm m}$ 48$\fs$4, 
Dec = $-$09$^{\circ}$ 20$'$ 51$\farcs$3 with a magnitude R $\sim$18 
(Peterson \& Price 2003; Price \& Peterson 2003). Martini et al. (2003a) 
measured the redshift of this Optical Transient (OT), $z$ = 1.52, based on 
the identification of Mg {\sc ii}, Fe {\sc ii} and Al {\sc ii} absorption.  
This redshift was confirmed by Rol et al. (2003b), who also reported the 
possible detection of a foreground absorption system at redshift $z$ = 
1.29, and by Fugazza et al. (2003a). Multicolor imaging and analysis of 
the host galaxy was reported by Gorosabel et al. (2005). {\it Chandra} 
observations, starting about 15 hours after the GRB, revealed a new, 
fading X--ray source at a position consistent with that of the optical 
transient, identified with the X-ray afterglow (Butler et al. 2005). Its 
temporal decrease followed a powerlaw with index $\alpha_{\rm X}$ = 1.5 
$\pm$ 0.1, and its 0.5--5 keV spectrum was well fitted by an absorbed 
power-law of photon index $\Gamma_{\rm X} = 2.0 \pm 0.2$. The N$_{\rm H}$ 
column density derived from the spectral fit, (5 $\pm$ 3) $\times$ 
10$^{20}$ cm$^{-2}$, is consistent with the Galactic foreground value 
(Dickey \& Lockman 1990).

We report here on the study of the optical afterglow emission of GRB 
030328. The paper is organized as follows: Sect. 2 describes the 
observations and data reduction; the photometric, spectroscopic and 
polarimetric results together with the multiwavelength spectrum are 
reported in Sect. 3; discussion and conclusions are presented in Sect. 4.

Most of the data presented in this paper were acquired, with programs led 
by the GRACE\footnote{GRB Afterglow Collaboration at ESO: see the web page 
{\tt http://www.gammaraybursts.org/grace/}} collaboration, at the Nordic 
Optical Telescope (NOT) in the Canary Islands (Spain) and at the European 
Southern Observatory (ESO) telescopes of Cerro Paranal and La Silla 
(Chile), starting $\sim$11 hours after the high-energy prompt event. 
Observations acquired with the 40-inch SSO telescope, the 1m Sampurnanand 
telescope at the Aryabhatta Research Institute of Observational Sciences 
(ARIES) in Naini Tal (India) and with the 1m US Naval Observatory 
Flagstaff Station (NOFS) telescope (Arizona, USA) are also included in 
this paper, along with the analysis of public VLT optical polarimetry 
data.

We remark that, although GRB 030328 was a bright event, it was not 
monitored intensively after the first hours, because it occurred the day 
before GRB 030329, the extremely bright, low redshift ($z$ = 0.168; 
Greiner et al. 2003b) event triggered by {\it HETE-2} (Vanderspek et al. 
2003) and unambiguously associated with a supernova (Stanek et al. 2003; 
Hjorth et al. 2003), which caught all the observers' attention over 
subsequent weeks.

Throughout this paper we will assume a cosmology with $H_0$ = 65 km 
s$^{-1}$ Mpc$^{-1}$; $\Omega_{\Lambda}$ = 0.7; $\Omega_{\rm m}$ = 0.3; 
also, when not otherwise indicated, errors and upper limits will be given 
at 1$\sigma$ and 3$\sigma$ confidence levels, respectively.


\section{Observations and data reduction}


\begin{figure*}[th!]
\vspace{-.5cm}
\psfig{file=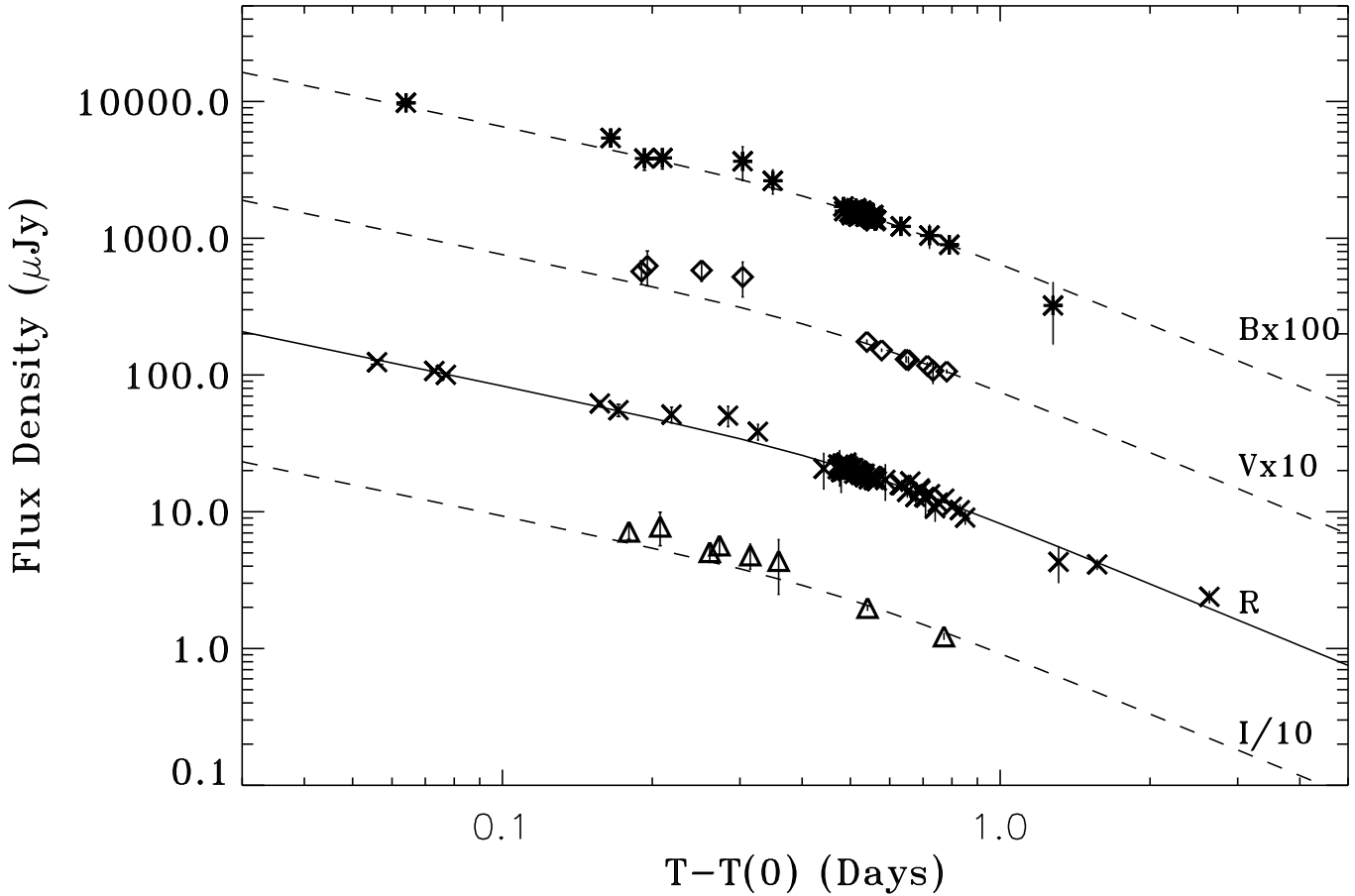,width=18.8cm}
\vspace{-.5cm}
\caption{$BVRI$ light curves of the GRB 030328 afterglow. Upper limits 
were not plotted. For clarity, $B$, $V$ and $I$ light curves are shifted 
by the factors indicated in the Figure. Here, T(0) corresponds to the GRB 
030328 trigger time (2003 Mar 28.4729 UT). The data points reported here 
are from this paper and from Bartolini et al. (2003), Burenin et al. 
(2003), Fugazza et al. (2003b), Gal-Yam et al. (2003), Garnavich et al. 
(2003), Ibrahimov et al. (2003) and Martini et al. (2003b). For these 
measurements, when no error was reported, a 0.3 mag uncertainty was 
assumed. Data are corrected for the underlying host galaxy contribution 
and for Galactic absorption. The solid line represents the best-fit of the 
$R$-band light curve using a broken powerlaw with indices $\alpha_1$ = 
0.76, $\alpha_2$ = 1.50, and break time $t_b$ = 0.48 d after the GRB. This 
same model, with unchanged model parameters except for the normalization, 
is overplotted as a dashed curve onto the $BVI$ data points and describes 
well the decay behaviour of these light curves.}
\end{figure*}


\subsection{Photometry}

Imaging in the Johnson-Cousins $UBVRI$ and Gunn $i$ filters was performed 
between March 28 and March 31 at the 2.5m NOT telescope in La Palma with 
the ALFOSC instrument. This is equipped with a 2048$\times$2048 pixel 
Loral CCD, giving a field of view of 6$\farcm$4$\times$6$\farcm$4 and an 
image scale of 0$\farcs$188 pix$^{-1}$.

Optical $BVRI$ data were also obtained on 2003 Mar 29 using the 2.2m 
ESO/MPG and 1.54m Danish telescopes at La Silla. The Wide Field Imager 
(WFI), permanently mounted on the 2.2m ESO/MPG telescope, is a focal 
reducer-type camera, compraising a mosaic of 8 2048$\times$4096 pixel 
CCDs. The WFI provides a field of view of 34$'$$\times$33$'$ and an image 
scale of 0$\farcs$238 pix$^{-1}$. The DFOSC is equipped with a 
2048$\times$2048 pixels CCD covering a 13$\farcm$7$\times$13$\farcm$7 
field, with a spatial resolution of 0$\farcs$39 pix$^{-1}$. On the same 
night, $V$-band data were acquired with VLT-{\it Antu} plus FORS1 at Cerro 
Paranal. FORS1 is equipped with a 2048$\times$2048 pixel Tektronix CCD 
which covers a 6$\farcm$8$\times$6$\farcm$8 field in the standard 
resolution imaging mode with a scale of 0$\farcs$2 pix$^{-1}$.

$UBVRI$ frames were also acquired with the 1m NOFS telescope. This 
telescope carries a 1024$\times$1024 pixel Tektronix CCD which can image a 
field of 12$'$$\times$12$'$ with a scale of 0$\farcs$68 pix$^{-1}$.

Earlier observations in the $B$ and $R$ filters were obtained with the 
40-inch SSO telescope and with the 1m ARIES telescope. The 40-inch 
telescope was equipped with a WFI, consisting of 8 2048$\times$4096 pixel 
CCDs, which cover a field of view of 52$'$$\times$52$'$ with an image 
scale of 0$\farcs$38 pix$^{-1}$. The 1m ARIES telescope carries a 
2048$\times$2048 pixel CCD, with a 13$'$$\times$13$'$ field of view; in 
its standard 2$\times$2 binning mode it has a spatial resolution of 
0$\farcs$76 pix$^{-1}$.

Optical frames were bias-subtracted and flat-fielded with the standard 
cleaning procedure. We chose standard Point Spread Function (PSF) fitting 
to perform photometry on the images, and used the {\sl DAOPHOT II}~image 
data analysis package PSF-fitting algorithm (Stetson 1987) running within 
MIDAS\footnote{MIDAS (Munich Image Data Analysis System) is developed, 
distributed and maintained by ESO and is available at {\tt 
http://www.eso.org/projects/esomidas/}}. A two-dimensional Gaussian 
profile with two free parameters (the half width at half maxima along the 
$x$ and $y$ coordinates of each frame) was modeled on at least 5 
unsaturated bright stars in each image.

The $UBVRI$ zero-point calibration was performed using the photometry by 
Henden (2003). We then selected stars of varoius brightness in the GRB 
030328 field and used them to determine the $UBVRI$ magnitudes of the OT. 
No color-term correction was computed as we chose stars with color indices 
as close as possible to those of the OT. The errors associated with the 
measurements reported in Table 1 account for both statistical 
uncertainties obtained with the standard PSF-fitting procedure and 
systematic errors of the magnitude calibration. The single Gunn $i$ image 
acquired with the NOT was calibrated using the $I$-band secondary 
standards, given that the widths, the reference wavelengths and the flux 
density normalizations of the two filters are very similar (Fukugita et 
al. 1995). However, in order to account for small differences between the 
filters, we added in quadrature a 3\% uncertainty to the magnitude error 
obtained from the Gunn $i$ observation.

All data reported in this paper are listed in Table 1 and are shown in 
Fig. 2, where we also report those available in the GCN Circulars' 
archive\footnote{{\tt http://gcn.gsfc.nasa.gov/gcn/gcn3\_archive.html}}, 
rescaled to the photometry of Henden (2003) when needed. For the cases in 
which no error was reported, a 0.3 mag uncertainty was assumed. We did not 
include the single $R$-band data point of Rumyantsev et al. (2003) because 
it deviates by about 4$\sigma$ from the more precise and 
nearly-simultaneous data secured at the 2.5m NOT.

The results presented here and obtained from the 40-inch SSO telescope 
data supersede the preliminary ones of Peterson \& Price (2003) and Price 
\& Peterson (2003).
   
The optical data of Fig. 2 were corrected for the Galactic foreground 
reddening assuming $E(B-V)$ = 0.047 mag (Schlegel et al. 1998): by 
applying the relation of Cardelli et al. (1989), we derived $A_B$ = 0.19 
mag, $A_V$ = 0.15 mag, $A_R$ = 0.12 mag and $A_I$ = 0.09 mag. Next, they 
were converted into flux densities assuming the normalizations of Fukugita 
et al. (1995). The host galaxy emission in the $BVRI$ bands was computed 
from the data of Gorosabel et al. (2005) and subtracted from our optical 
data set.

\subsection{VLT spectroscopy}


\begin{figure*} 
\vspace{-5cm}
\psfig{file=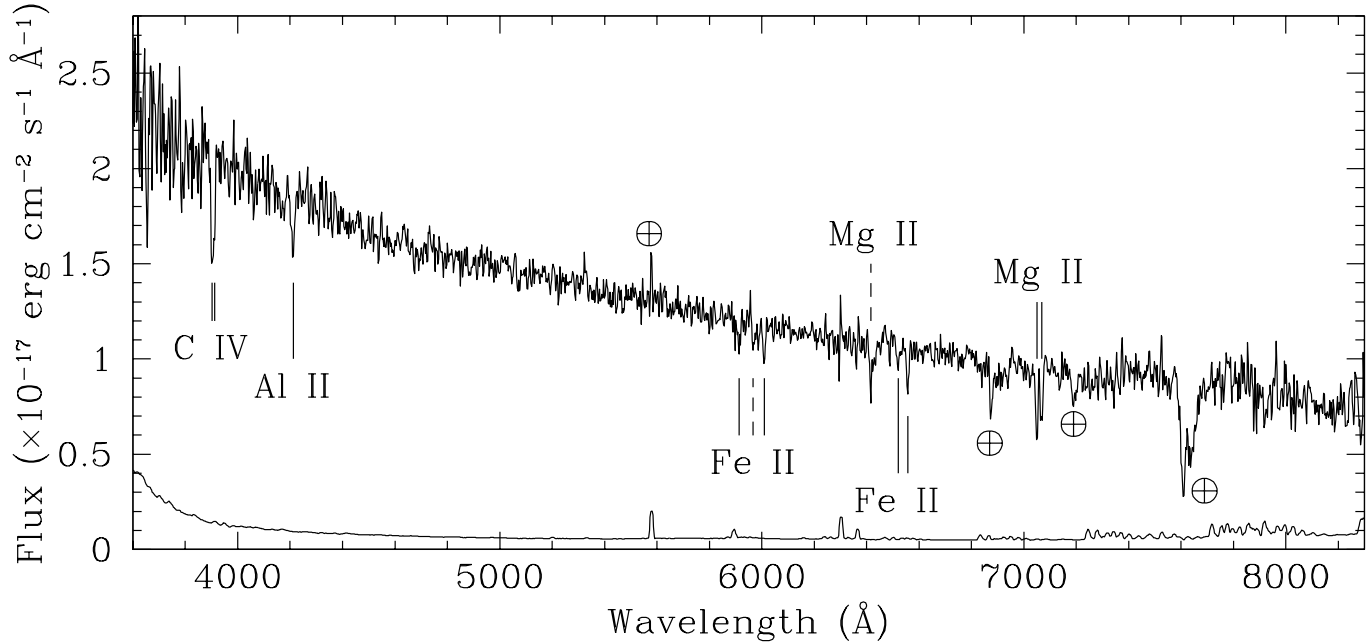,width=18.5cm,angle=270} 
\caption{Optical spectrum of the afterglow of GRB 030328. A number of 
rest-frame ultraviolet metal absorption lines are detected (see also Table 
2). To explain all of them, at least two systems at redshift $z$ = 1.5216 
and $z$ = 1.295 are required. The system at the highest redshift (the 
lines of which are indicated with solid hyphens) likely originates in the 
host galaxy, whereas the lower one is produced by a foreground absorber; 
the lines of the latter are marked with dashed hyphens. The symbol 
$\oplus$ indicates atmospheric and telluric features. The solid line 
towards the bottom of the Figure represents the 1$\sigma$ Poisson error 
spectrum.}
\end{figure*}


As reported in Table 1, a series of six 10-min spectra was obtained 
starting on 2003 Mar 29.060 UT (0.59 d after the GRB) at ESO-Paranal using 
VLT-{\it Antu} plus FORS1 under fairly good weather conditions, with 
seeing 0$\farcs$9 and airmass 1.18--1.41. The 300V Grism was used with a 
nominal spectral coverage of 3600--8000 \AA. The slit width was 1$\arcsec$ 
and the optical spectral dispersion was 2$\farcs$7 \AA/pixel 
($\lambda/\Delta\lambda$ = 440).

All spectra have been reduced in the IRAF\footnote{available at {\tt 
http://iraf.noao.edu}} environment with the {\tt kpnoslit} package. After 
overscan subtraction and flat-fielding, cosmic rays were removed from the 
images using the L.A. Cosmic routine written by van Dokkum (2001). The 
spectra were then optimally extracted for each 2-D image separately, with 
an extraction radius of about 2\arcsec. The dispersion correction was 
applied, again to each spectrum separately, using an HeNeAr lamp spectrum 
that was taken in the morning after the science observations. The typical 
formal error in the wavelength calibration fit was 0.2 \AA. The separate 
dispersion-corrected spectra were averaged, and the corresponding Poisson 
error spectra, calculated by the {\tt apall} task, were quadratically 
averaged.

The flux calibration was performed using observations of the standard LTT 
7379 (Hamuy et al. 1992, 1994) carried out during the same night as those 
of the GRB OT spectrum.

We applied a slit-loss correction by fitting a Gaussian profile along the 
spatial direction of the averaged 2-D spectra, every 4 pixels across the 
entire dispersion axis (i.e. summing 4 columns before performing the fit). 
The resulting Gaussian full width at half maximum (FWHM) was then compared 
to the slit width in order to obtain the slit loss (i.e. the fraction of 
the surface underneath the Gaussian fit that is outside the slit width) 
along the dispersion axis. The slit loss profile was then fitted with a 
polynomial to correct the spectra. Note that this does not correct for any 
colour-dependent slit losses, but FORS1 has a linear atmospheric 
dispersion compensator (LADC) in the light path which minimizes any 
colour-differential slit losses up to a zenith distance of 45\degr.

\subsection{VLT polarimetry}

Linear polarimetry $V$-band observations (see Table 1 for the observation 
log) were acquired starting on 2003 Mar 29.130 UT at VLT-{\it Antu} plus 
FORS1 under an average seeing of $\sim$1$\arcsec$. Data were obtained, 
between 0.66 and 0.88 d after the GRB, using a Wollaston prism and a 
half-wavelength, phase-retarder plate. The Wollaston prism separates the 
incident light into ordinary and extraordinary components, while the 
phase-retarder plate determines which of the Stokes parameters is measured 
(U or Q). For each image, a mask producing 22$''$-wide parallel strips was 
used to avoid so as overlap of the two components. The observation 
therefore consisted of four exposures centered at the position of the OT, 
with the phase-retarder plate at 0$^{\circ}$, 22$\fdeg$5, 45$^{\circ}$ and 
67$\fdeg$5. Each angle was imaged 5 times cyclically with an exposure time 
of 780 s, for a total exposure time of 15600 s over the 4 angles. Image 
reduction and analysis was performed as described in Sect. 2.1 for the 
optical photometry.

In order to evaluate the Stokes U and Q parameters of the OT emission, and 
thus its polarization percentage $P$ and the position angle of the 
electric field vector $\theta$, we applied the method described by Di 
Serego Alighieri (1997). Here $P$ and $\theta$ are obtained by using the 
relation $S(\phi) = P~ cos~2(\theta-\phi)$, in which $S$ depends on the 
ratio between the fluxes of the ordinary and extraordinary components of 
the incident beam, and $\phi$ corresponds to the prism rotation angle. The 
values of $P$ and $\theta$ are calculated by fitting the above relation to 
the measurements obtained in correspondence with each rotation angle. 
Moreover, using this formalism, one obtains the two Stokes parameters Q 
and U from the values of $S(0^{\circ})$ and $S(45^{\circ})$, respectively.

No significant variation among the values of the U and Q parameters of 
individual field stars was noted, therefore we computed the average values 
of these quantities and subtracted them from the corresponding parameters 
of the OT to remove the instrumental and (local) interstellar 
polarization. We also checked that the polarization of the selected field 
stars did not systematically vary across the CCD or with the star 
magnitudes.

\subsection{{\it Chandra} X--ray data}

In order to determine a precise astrometric position for the X--ray 
afterglow of GRB 030328, we retrieved the public {\it Chandra}/ACIS-S 
observation\footnote{available at \\ {\tt http://heasarc.gsfc.nasa.gov}} 
of this source (see Butler et al. 2005 for details). The packages 
CIAO\footnote{available at {\tt http://cxc.harvard.edu/ciao/}} v3.2.2 and 
CALDB\footnote{available at {\tt http://cxc.harvard.edu/caldb/}} v3.1.0 
were used for the data reduction. The aspect-solution 90\% confidence 
level error radius of 0$\farcs$6 was assumed for the positions of the 
X--ray sources determined from this observation.


\section{Results}

\subsection{Optical and X--ray astrometry}

The OT (with $R$ = 21.17 mag at 0.80 d after the GRB; see Table 1) was 
clearly detected in the WFI $R$-band image (see Fig. 1). An astrometric 
solution based on 50 USNO-A2.0\footnote{The USNO-A2.0 catalogue is 
available at \\ {\tt http://archive.eso.org/skycat/servers/usnoa}} 
reference stars in the WFI $R$ image, taken on 2003 Mar 29.273 UT, yields 
for the OT the coordinates (J2000) RA = 12$^{\rm h}$ 10$^{\rm m}$ 
48$\fs$37, Dec = $-$09$^{\circ}$ 20$'$ 51$\farcs$39. The internal error of 
the optical position is 0$\farcs$238, which has to be added to the 
systematic error of the USNO catalogue (0$\farcs$25 according to Assafin 
et al. 2001 and Deutsch 1999). The final astrometric error is thus 
0$\farcs$35.

The position of the X--ray source corresponding to the GRB 030328 
afterglow in the {\it Chandra} data was obtained using the {\tt 
celldetect} command. Only a single bright source is detected at the center 
of the ACIS-S field of view. This does not allow us to tie the internal 
{\it Chandra} astrometry to that of optical and/or near-infrared 
catalogues in order to improve the positional uncertainty afforded by the 
ACIS-S X--ray data.

The {\it Chandra}/ACIS-S astrometry we obtained for the X--ray afterglow 
of GRB 030328 is the following (J2000): RA = 12$^{\rm h}$ 10$^{\rm m}$ 
48$\fs$384, Dec = $-$09$^{\circ}$ 20$'$ 51$\farcs$71, with an error of 
0$\farcs$6 on both coordinates. This position lies 0$\farcs$38 from, and 
is therefore fully consistent with, that reported above for the GRB 030328 
OT.

\subsection{Optical light curves}

In Fig. 2 we plot our photometric measurements together with those 
reported in the GCN circulars. As this figure shows, the best-sampled 
curve is the $R$-band, so we will first consider this for the light curve 
decay fitting.

First we modeled the $R$-band light curve using a single powerlaw with 
index $\alpha$ = 0.94 $\pm$ 0.01 ($\chi^2/dof$ = 145.8/81, where $dof$ 
means `degrees of freedom'), but this model did not satisfactorily 
describe the data, with the points corresponding to the early ($<$0.3 d 
after the GRB) and late ($>$1 d after the GRB) observations lying 
systematically below the best-fit curve. The fit improves significantly by 
using a smoothly broken powerlaw (see Beuermann et al. 1999). The best fit 
(with $\chi^2/dof$ = 0.4/79), plotted in Fig. 2, is provided by the 
temporal indices $\alpha_1 = 0.76 \pm 0.03$ and $\alpha_2 = 1.50 \pm 0.07$ 
before and after a break occurring at $t_b = 0.48 \pm 0.03$ d from the GRB 
trigger, and with $s = 4.0 \pm 1.5$ the parameter describing the slope 
change rapidity.

The above best fit also describes the second well-sampled band, i.e., the 
$B$ one. Concerning the other optical bands, the number of available data 
points is not enough to allow a meaningful fitting; however, the best-fit 
obtained for the $R$-band light curve is again fully consistent with the 
decay trend in the $V$ and $I$ bands (see Fig. 2). This means that the 
decay of the OT can be considered as achromatic within the uncertainties 
of the $R$-band light curve best-fit parameters.

We also note that there is a marginal presence of a deviation from the 
best-fit curve, in the form of an increase in brightness, starting 
$\sim$0.2 d after the trigger and lasting $\sim$0.15 d. This variation is, 
moreover, apparent in all optical bands and is consistent with being 
achromatic within our uncertainties.

The optical colors of the OT of GRB 030328 fall in the loci populated by 
GRB afterglows in the color-color diagrams as illustrated by \v{S}imon et 
al. (2001).

\subsection{Spectroscopy}

Figure 3 shows the spectrum of the GRB 030328 OT. As already remarked in 
Sect. 2.2, the resolving power is roughly 440 at 5900 \AA, corresponding 
to about 700 km s$^{-1}$, or 13.4 \AA. Table 2 lists the significant 
(3$\sigma$) lines that we detect, their identification, and rest-frame 
equivalent width (EW) W$_r$. Most of the significant features can be 
identified with metal absorption lines in a system at a redshift of $z$ = 
1.5216 $\pm$ 0.0006. These lines are likely associated with the 
circumburst gas or interstellar medium (ISM) in the GRB host galaxy. A 
lower redshift absorption system at $z$ = 1.295 $\pm$ 0.001 is also found: 
for it, only two lines can be identified (Fe {\sc ii} $\lambda$2600 and 
the unresolved Mg {\sc ii} $\lambda$$\lambda$2796,2803 doublet). Its 
detection indicates the presence of a foreground absorber.

We estimate metal column densities using the observed EW (Table~2), as 
described in Savaglio \& Fall (2004). Stringent constraints cannot be 
derived because the resolution of the spectrum does not allow the 
detection of the many weak absorption lines. For Fe {\sc ii} (see the 
corresponding Curve of Growth in Fig. 4), we derive $\log N_{\rm Fe\, II} 
= 14.3^{+0.6}_{-0.2}$ and an effective Doppler parameter $b$ =$ 
31^{+13}_{-10}$ km s$^{-1}$. The column density of Mg {\sc ii} is not well 
constrained, but is likely in the range $\log N_{\rm Mg\, II} = 
13.9-15.3$. For Si {\sc ii} and Al {\sc ii} we derive $\log N_{\rm Al\, 
II} < 16.0$ and $\log N_{\rm Si\, II} < 16.0$ assuming a conservative 
value $b>25$ km s$^{-1}$ for the Doppler parameter.


\begin{figure}
\psfig{file=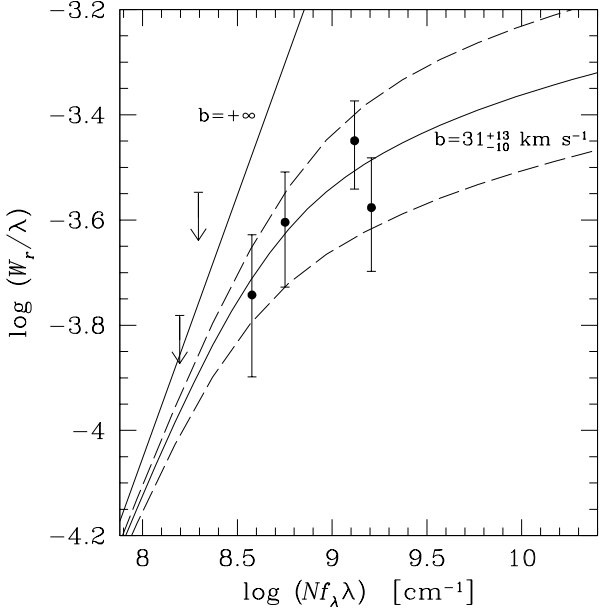,width=8.5cm}
\caption{Curve of Growth for the Fe {\sc ii} absorption lines detected in 
the GRB 030328 OT spectrum at redshift $z$ = 1.5216. The curved solid line 
represent the best fit, whereas the straight solid line marks the $b$ = 
+$\infty$ case. Dashed lines indicate the 1-$\sigma$ borders of the 
best-fit curve.}
\end{figure}


As also suggested from the X--ray spectral fitting (Butler et al. 2005), 
weak absorption is associated with this GRB environment; thus, we do not 
expect dust in the neutral gas to obscure significantly the optical 
afterglow. From the Fe {\sc ii} column density, we estimate $A_V < 0.1$ 
mag (Savaglio et al. 2003; Savaglio \& Fall 2004). The dust extinction in 
the intervening system at $z=1.295$ is even more negligible.

\subsection{Polarimetry}

In Fig. 5 the polarimetric data analysis results are shown. The intrinsic 
nature of the OT $V$-band polarization is supported by the relative 
location of the OT with respect to that of field stars in the U vs. Q
plot (Fig. 5, upper panel).


\begin{figure}[t!]
\psfig{file=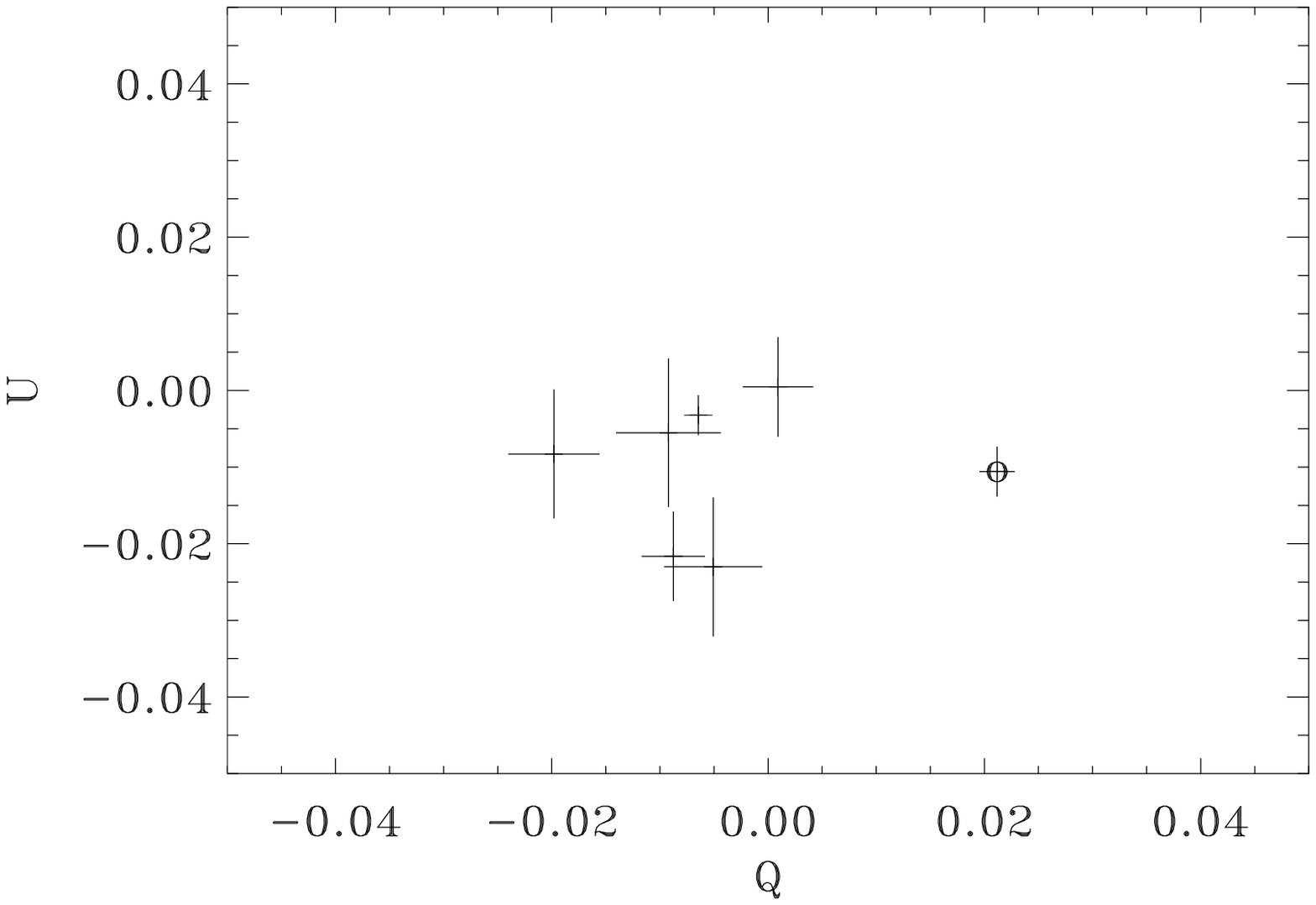,width=8.5cm}
\psfig{file=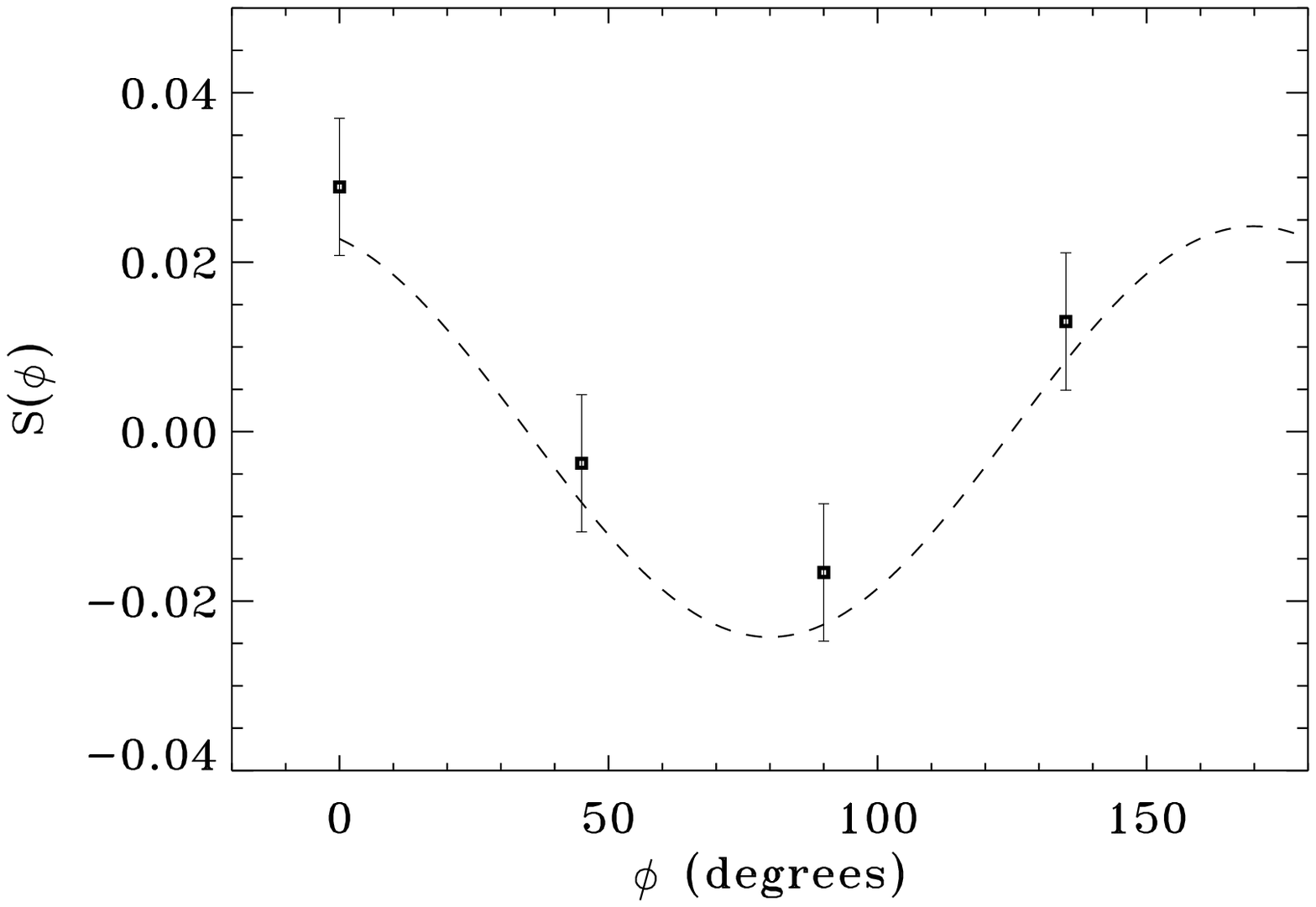,width=8.5cm}
\caption{(Upper panel) positions of field stars and of the OT (marked 
with the open circle) in the plane of Q and U parameters not corrected 
for spurious (instrumental plus field) polarization. The OT is clearly 
separated from the region occupied by the field stars: this indicates 
that it has net intrinsic polarization. (Lower panel) cosine fit of our
$V$-band polarimetric data. The best-fit yields $P = (2.4 \pm 0.6)$ \% 
and $\theta = 170^{\circ} \pm 7^{\circ}$.}
\end{figure}


After correcting for spurious field polarization, we found Q$_{\rm OT} = 
0.029 \pm 0.008$ and U$_{\rm OT} = -0.004 \pm 0.008$. The fit of the data 
with the relation of Di Serego Alighieri (1997; see Fig. 5, lower panel) 
yielded for the OT a linear polarization percentage of $P = (2.4 \pm 0.6)$ 
\% and a polarization angle $\theta = 170^{\circ} \pm 7^{\circ}$, 
corrected for the polarization bias (Wardle \& Kronberg 1974). This latter 
correction is introduced because $P$ is a positive definite quantity, and 
thus at low signal-to-noise (S/N) polarization levels the distribution 
function of $P$ is no longer normal but becomes skewed, which causes an 
overestimate of the real value of $P$ (Simmons \& Stewart 1985).

The polarimetric observations covered the time interval 0.66--0.88 d after 
the GRB, during which the total flux of the OT decreased by $\sim$0.5 mag 
assuming the afterglow temporal decay as modeled in Sect. 3.2. In order to 
check whether this could have affected our measurements, we also 
separately considered each of the 5 single polarimetry cycles (see Table 
3). Although with lower S/N, $P$ and $\theta$ are consistent with being 
constant across the whole polarimetric observation run (i.e., over 
$\sim$5.3 hours). This justifies our choice to consider the total 
polarimetric images, summed over each angle, for the determination of the 
$P$ and $\theta$ values in order to increase the final S/N of the 
measurement.

The use of standard polarization equations (e.g., Ramaprakash 1998) leads 
to results which are in good agreement, within the uncertainties, with 
those obtained from the method described above.

Once the correction for spurious polarization, induced by the Galactic 
ISM, has been taken into account, one may be concerned with the presence 
of further spurious polarization effects local to the GRB host. However, 
regarding possible additional local dust absorption within the host, we 
have shown in Sect. 3.3 that we do not find any evidence for this, based 
on the analysis of the optical spectrum of the GRB 030328 afterglow. 
Moreover, the contribution of the host galaxy to the total (OT+host) light 
is just a few percent; therefore we do not expect that the host emission 
significantly contaminated our polarization measurement. Therefore we 
conclude that the measured OT polarization is intrinsic.

\subsection{The broadband spectrum of the afterglow}


\begin{figure}[t!]
\psfig{file=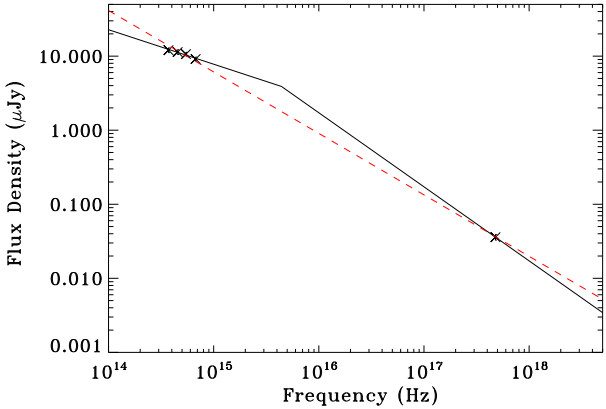,width=8.5cm}
\caption{Broadband spectrum at the epoch corresponding to 0.78 d after the 
GRB. The dashed line corresponds to a fit of optical and X--ray bands 
using a single powerlaw with a spectral index of 0.8; the broken solid 
line indicates that a good fit is also obtained assuming the cooling 
frequency between optical and X--ray bands, at $5.9 \times 10^{15}$ Hz.}
\end{figure}


By using the available information, we have constructed the 
optical-to-X--ray SFD of the GRB 030328 afterglow at 0.78 days after the 
GRB, that is, the epoch with best broadband photometric coverage.

The optical flux densities at the wavelengths of the $BVRI$ bands have 
been derived by subtracting the host galaxy contribution at the epoch we 
selected. As in Sect. 2.1, we assumed $E(B-V) = 0.047$ to deredden the 
data and the normalizations given in Fukugita et al. (1995) to obtain the 
corresponding flux densities in the optical band. When needed, we rescaled 
the data to the corresponding reference epoch, 2003 Mar 29.253 UT, using 
the optical powerlaw decay with (post-break) index $\alpha_{\rm opt} = 
1.50 \pm 0.07$ as the SFD reference epoch is after $t_b$. We also rescaled 
to the SFD epoch the X--ray flux corresponding to the logarithmic center 
of the X--ray spectrum energy range (0.5--5 keV, see Butler et al. 2005): 
this was done using the powerlaw decay found by these authors for the 
X--ray afterglow temporal behaviour, $\alpha_{\rm X} = 1.5 \pm 0.1$.

As Fig. 6 shows, we fitted the data in two different ways. First we used a 
single powerlaw (dashed line) to fit the optical and the X--ray data, and 
we found a spectral index $\beta_{\rm X-opt}$ = 0.83 $\pm$ 0.01. However, 
when we independently fit the optical data with a powerlaw, we obtain a 
spectral slope $\beta_{\rm opt}$ = 0.47 $\pm$ 0.15, which is marginally 
flatter than the X--ray one ($\beta_{\rm X}$ = $\Gamma_{\rm X} -$ 1 = 1.0 
$\pm$ 0.2; Butler et al. 2005).

In this latter case, represented by the broken solid line in Fig. 6, the 
presence of a spectral steepening between optical and X--ray bands could 
be explained with the presence of a cooling frequency $\nu_{\rm c}$ in the 
framework of the synchrotron fireball model (Sari et al. 1998). Assuming a 
negligible host absorption (see Sect. 3.3) and using the optical and 
X--ray slopes above, we obtain $\nu_{\rm c}$ = $5.9 \times 10^{15}$ Hz.


\section{Discussion}

The afterglow of GRB 030328 was imaged in the optical for $\sim$2.5 days, 
starting a few hours after the explosion. Optical spectroscopy and 
polarimetry have also been performed. By combining these data with 
observations at other frequencies, we can compare the temporal and 
spectral behaviour of its broad-band afterglow with the fireball scenario, 
assuming a regime of adiabatic expansion (Sari et al. 1998, 1999).

The steepening of the $R$-band light curve around 0.5 days after the GRB 
suggests the presence of a jet break and makes the scenario of a spherical 
expansion less probable. Therefore, hereafter we will consider only a 
jet-collimated expansion in a homogeneous medium (Sari et al. 1999; Rhoads 
1999) or in a wind-shaped medium (Chevalier \& Li 2000).

The optical and X--ray spectral indices ($\beta_{\rm opt}$ = 0.47 $\pm$ 
0.15 and $\beta_{\rm X}$ = 1.0 $\pm$ 0.2) are marginally consistent with 
each other, therefore we cannot exclude that a single power-law with index 
$\beta_{\rm opt} = \beta_{\rm X} = 0.83 \pm 0.01$ is the best 
approximation of the optical-to-X-ray spectrum (see Fig. 6).

The first case we deal with is that in which the optical and X--ray bands 
have different spectral slopes. Assuming $\beta_{\rm X} = p/2$, we obtain 
$p = 2.0 \pm 0.4$ and $\beta_{\rm opt} = (p-1)/2 = 0.5 \pm 0.2$, which is 
fully consistent with the value we found from the broadband spectrum (see 
Sect. 3.5). Moreover, if we consider a fireball model consisting of a 
fixed-angle jet-collimated expansion in a homogeneous medium (M\'esz\'aros 
\& Rees 1999), we obtain $\alpha_1^{\rm model} = 3(p-1)/4 = 0.75 \pm 0.3$ 
and $\alpha_2^{\rm model} = 3(p-1)/4+3/4 = 1.5 \pm 0.3$ before and after 
the jet break, respectively. These values are consistent (within the 
errors) with the broken powerlaw decay indices of the $R$-band light 
curve. Moreover, the computed post-break decay index $\alpha_2^{\rm 
model}$ is consistent with the X--ray temporal slope as determined by 
Butler et al. (2005), as the {\it Chandra} observation starts after the 
optical jet break time.

In the second case we consider that the optical-to-X-ray spectrum is 
produced by the same emission component so that it is described by a 
single spectral slope, $\beta$ = 0.83. If the cooling frequency $\nu_c$ is 
below the optical band ($\nu_c < \nu_{opt}$), one would have $\beta_{\rm 
X} = \beta_{\rm opt} = p/2$, so we obtain for the electron energy 
distribution index $p$ the value 1.66$\pm$0.02 and a temporal decay index 
$\alpha < 1$ (see the relations of Dai \& Cheng 2001 for the cases in 
which $p<2$). The latter is however not consistent with the observed 
optical and X--ray temporal indices at the SFD time. Otherwise, if 
$\nu_{\rm c} > \nu_{\rm X}$, which implies $\beta_{\rm X} = \beta_{\rm 
opt} = (p-1)/2$, we obtain $p = 2.66\pm0.02$: also in this case, the value 
for $\alpha$ after the jet break time (expected to be numerically equal to 
$p$) is not consistent with our observations. Even assuming a fixed 
collimation angle, the inconsistency between the model and the 
observations persists.

Similarly, assuming a fireball expansion in a wind-shaped medium 
(Chevalier \& Li 2000) returns temporal decay values that are inconsistent 
with those observed. Thus, this option is not viable as a description of 
the GRB 030328 afterglow evolution.

We conclude that the favoured scenario is a jet-collimated expansion in a 
homogeneous medium with fixed opening angle. From the measured jet break 
time we can compute the jet opening angle value for GRB 030328 which is, 
following Sari et al. (1999), $\theta_{\rm jet} \sim 3\fdg2$.

In this hypothesis, considering a luminosity distance $d_L$ = 11.96 Gpc 
and a GRB fluence $S_{\gamma}$ = 3.0$\times$10$^{-5}$ erg cm$^{-2}$ 
(Villasenor et al. 2003), we find that $E_{\rm iso} = k \cdot (4\pi 
S_{\gamma} d_L^2)/(1+z) \sim 2.0 \times 10^{53}$ erg, where $k$ is the the 
bolometric correction factor, which we assume to be $\sim$1 (Bloom et al. 
2001). With the information above, we can estimate the value of the total 
energy corrected for the jet collimation as $E_\gamma \sim 3.1 \times 
10^{50}$ erg. 
This $E_\gamma$ value, when compared with the rest-frame peak energy 
$E_{\rm peak}$ of the $\nu F \nu$ spectrum of the prompt event (Barraud et 
al. 2004) is consistent within 1$\sigma$ with the `Ghirlanda relation' 
(Ghirlanda et al. 2004) which ties these two quantities together.

The presence of a bump during the first phases of the optical decay is not 
uncommon in GRB OTs (see e.g. GRB 000301c, Masetti et al. 2000; GRB 
021004, de Ugarte Postigo et al. 2005; GRB 050502a, Guidorzi et al. 2005). 
This behaviour has generally been interpreted as due to late energy 
injection produced by a refreshed shock (Kumar \& Piran 2000; Bj\"ornsson 
et al. 2004). This mechanism may explain the early bump we possibly see 
starting $\sim$0.2 d after the high-energy prompt event in the OT light 
curves of GRB 030328, although a detailed analysis is hindered by the low 
S/N of the feature.

Concerning optical spectroscopy, we note that the absorption system at $z$ 
= 1.5216, likely associated with the host of GRB 030328 (Fig. 3), is one 
of the weakest ever observed in a GRB optical transient. The rest-frame EW 
of the Fe {\sc ii} $\lambda2600$ line (Table~2) is only 0.9 \AA\, and the 
total EW of the Mg {\sc ii} doublet is 2.9 \AA. To put this in context, 
all the Mg {\sc ii} and Fe {\sc ii} $\lambda2600$ absorptions detected so 
far in GRB afterglows ($\sim$10 objects) are larger than 3 \AA\ and 1 \AA, 
respectively, with the exception of GRBs 021004 (Fiore et al. 2005; 
Mirabal et al. 2003) and 990510 (Vreeswijk et al. 2001). Indeed 
GRB~990510, studied by Vreeswijk et al. (2001), has Mg {\sc ii} and Fe 
{\sc ii} $\lambda2600$ EWs very similar to those of GRB~030328. 
Incidentally, the spectral slope of the optical afterglow in these two GRB 
OTs are also similar: assuming $F_\lambda\propto\lambda^{-\delta}$, one 
finds $\delta\simeq1.5$ in both cases.

We also note that the GRB~030328 host may be a GRB-Damped Lyman-$\alpha$ 
Absorber (GRB-DLA; Vreeswijk et al. 2004): indeed, in DLA galaxies it was 
found that Fe {\sc ii} $\lambda$2600 and Mg {\sc ii} $\lambda$2796 EWs are 
larger than 0.5 \AA, as is seen in the GRB~030328 OT spectrum.

The detection of net polarization may imply that some kind of asymmetry is 
present in the GRB 030328 afterglow emission. The most natural explanation 
is the presence of a jet geometry, which is thought to produce the 
polarization observed in GRB OTs (Sari 1999; Ghisellini \& Lazzati 1999). 
The fact that we detect optical polarization around the jet break time, 
and that it does not seem to vary in percentage and angle during our five 
polarimetric subsets acquired over a time span of $\sim$5.3 hours, argues 
against the possibility that it is produced in a homogeneous jet. Indeed, 
in this case two polarization peaks are expected, separated by a moment of 
null polarization (roughly coincident with the break time of the total 
light curve) over which the polarization angle changes by 90$^{\circ}$ 
(see e.g. Ghisellini \& Lazzati 1999; Lazzati et al. 2003; Rossi et al. 
2004). So, we would expect that $P$ vanishes at $t_b$ and that it and 
$\theta$ rapidly change (on timescales of tens of minutes; see Fig. 5 of 
Lazzati et al. 2003) across that epoch; however, no hint for both facts is 
found in the case of the GRB 030328 OT (see Table 3). Rather, the presence 
of net polarization and the constancy of $P$ and $\theta$ are expected 
around the jet break time if a structured jet is present (Rossi et al. 
2004): thus, we suggest that the collimated emission of GRB 030328 has a 
structured distribution of the emissivity across the jet aperture.

\begin{acknowledgements}
We thank the anonymous referee for several useful comments which helped us 
to improve the paper. We also thank the ESO staff astronomers for their 
help and efforts in obtaining the observations presented in this paper, 
and J. Greiner for useful comments on an earlier version of this 
manuscript. We are grateful to John Stephen for the language editing 
of the final version of the manuscript. We acknowledge Scott Barthelmy for 
maintaining the GRB Coordinates Network (GCN) and BACODINE services. This 
research has made use of NASA's Astrophysics Data System and HEASARC 
high-energy missions archive, and of data retrieved from the ESO/ST-ECF 
Science Archive. This work has been partially supported by the EU RTN 
Contract HPRN-CT-2002-00294.
\end{acknowledgements}



\begin{table*}[t!]
\caption[]{Log of the photometric, spectroscopic and polarimetric 
observations presented in this paper. No corrections for foreground 
Galactic absorption and for the underlying host contribution were applied 
to the reported magnitudes. Magnitude errors are at 1$\sigma$ confidence 
level, whereas upper limits are at 3$\sigma$ confidence level.}
\begin{center} 
\begin{tabular}{cccccc} 
\hline
\hline
\noalign{\smallskip}
Mid-exposure & Telescope & Filter   & Exposure & Seeing or & Magnitude \\
 time (UT)   &           & or grism & time (s) & slit width ($''$) & \\

\noalign{\smallskip}
\hline
\noalign{\smallskip}
\multicolumn{6}{c}{Photometry} \\
\noalign{\smallskip}
\hline
\noalign{\smallskip}

 2003 Mar 29.0109 & 2.5m NOT     & $U$ & 150 & 1.0 & 20.68$\pm$0.09 \\
 2003 Mar 29.4435 & 1.0m NOFS    & $U$ & 4$\times$600 & 5.4 & $>$20.2 \\
  &  &  &  & \\
 2003 Mar 28.5375 & 40-inch SSO  & $B$ & 600 & 2.6 & 19.22$\pm$0.04 \\
 2003 Mar 28.6386 & 40-inch SSO  & $B$ & 600 & 3.2 & 19.86$\pm$0.08 \\
 2003 Mar 28.6659 & 1.0m ARIES   & $B$ & 2$\times$300 & 2.3 & 20.23$\pm$0.19 \\
 2003 Mar 28.6826 & 1.0m ARIES   & $B$ & 300 & 2.6 & 20.22$\pm$0.18 \\
 2003 Mar 28.7769 & 1.0m ARIES   & $B$ & 300 & 2.0 & 20.28$\pm$0.29 \\
 2003 Mar 28.8223 & 1.0m ARIES   & $B$ & 2$\times$200 & 1.9 & 20.63$\pm$0.21 \\
 2003 Mar 28.9574 & 2.5m NOT     & $B$ &  60 & 1.0 & 21.09$\pm$0.04 \\
 2003 Mar 28.9592 & 2.5m NOT     & $B$ &  60 & 1.0 & 21.17$\pm$0.05 \\
 2003 Mar 28.9712 & 2.5m NOT     & $B$ &  80 & 1.0 & 21.19$\pm$0.04 \\
 2003 Mar 28.9738 & 2.5m NOT     & $B$ &  80 & 1.0 & 21.16$\pm$0.04 \\
 2003 Mar 28.9764 & 2.5m NOT     & $B$ &  90 & 1.0 & 21.24$\pm$0.05 \\
 2003 Mar 28.9789 & 2.5m NOT     & $B$ &  90 & 1.0 & 21.15$\pm$0.05 \\
 2003 Mar 28.9814 & 2.5m NOT     & $B$ &  90 & 1.0 & 21.19$\pm$0.04 \\
 2003 Mar 28.9839 & 2.5m NOT     & $B$ &  90 & 1.0 & 21.23$\pm$0.07 \\
 2003 Mar 28.9864 & 2.5m NOT     & $B$ &  90 & 1.0 & 21.12$\pm$0.04 \\
 2003 Mar 28.9892 & 2.5m NOT     & $B$ &  90 & 1.0 & 21.16$\pm$0.05 \\
 2003 Mar 28.9917 & 2.5m NOT     & $B$ &  90 & 1.0 & 21.27$\pm$0.04 \\
 2003 Mar 28.9942 & 2.5m NOT     & $B$ &  90 & 1.0 & 21.24$\pm$0.04 \\
 2003 Mar 28.9966 & 2.5m NOT     & $B$ &  90 & 1.0 & 21.25$\pm$0.04 \\
 2003 Mar 28.9991 & 2.5m NOT     & $B$ &  90 & 1.0 & 21.16$\pm$0.04 \\
 2003 Mar 29.0016 & 2.5m NOT     & $B$ &  90 & 1.0 & 21.22$\pm$0.04 \\
 2003 Mar 29.0041 & 2.5m NOT     & $B$ &  90 & 1.0 & 21.21$\pm$0.04 \\
 2003 Mar 29.0066 & 2.5m NOT     & $B$ &  90 & 1.0 & 21.20$\pm$0.04 \\
 2003 Mar 29.0093 & 2.5m NOT     & $B$ &  90 & 1.0 & 21.27$\pm$0.04 \\
 2003 Mar 29.0215 & 2.5m NOT     & $B$ &  90 & 1.0 & 21.28$\pm$0.04 \\
 2003 Mar 29.0240 & 2.5m NOT     & $B$ &  90 & 1.0 & 21.26$\pm$0.04 \\
 2003 Mar 29.0265 & 2.5m NOT     & $B$ &  90 & 1.0 & 21.31$\pm$0.04 \\
 2003 Mar 29.0291 & 2.5m NOT     & $B$ &  90 & 1.0 & 21.24$\pm$0.04 \\
 2003 Mar 29.0316 & 2.5m NOT     & $B$ &  90 & 1.0 & 21.31$\pm$0.04 \\
 2003 Mar 29.0341 & 2.5m NOT     & $B$ &  90 & 1.0 & 21.35$\pm$0.04 \\





 2003 Mar 29.0365 & 2.5m NOT     & $B$ &  90 & 1.0 & 21.33$\pm$0.04 \\
 2003 Mar 29.1045 & 2.5m NOT     & $B$ & 120 & 0.9 & 21.44$\pm$0.04 \\
 2003 Mar 29.2635 & 2.2m ESO/MPG & $B$ & 600 & 0.8 & 21.76$\pm$0.03 \\
 2003 Mar 29.3065 & 1.0m NOFS    & $B$ & 7$\times$1200 & 5.3 & 21.6$\pm$0.2 \\
  &  &  &  & \\
 2003 Mar 28.6633 & 1.0m ARIES   & $V$ & 300 & 2.1 & 19.63$\pm$0.21 \\
 2003 Mar 28.6685 & 1.0m ARIES   & $V$ & 300 & 2.7 & 19.53$\pm$0.30 \\
 2003 Mar 28.7243 & 1.0m ARIES   & $V$ & 300 & 3.0 & 19.61$\pm$0.18 \\
 2003 Mar 28.7769 & 1.0m ARIES   & $V$ & 300 & 2.0 & 19.73$\pm$0.30 \\
 2003 Mar 29.0127 & 2.5m NOT     & $V$ & 120 & 1.0 & 20.89$\pm$0.03 \\
 2003 Mar 29.0533 & VLT-$Antu$   & $V$ &  60 & 1.0 & 21.04$\pm$0.02 \\ 
 2003 Mar 29.1200 & VLT-$Antu$   & $V$ & 120 & 1.1 & 21.20$\pm$0.05 \\ 
 2003 Mar 29.1272 & VLT-$Antu$   & $V$ & 120 & 1.1 & 21.21$\pm$0.05 \\ 
 2003 Mar 29.1866 & VLT-$Antu$   & $V$ & 120 & 0.9 & 21.31$\pm$0.05 \\ 
 2003 Mar 29.2542 & 2.2m ESO/MPG & $V$ & 600 & 0.7 & 21.41$\pm$0.03 \\ 
 2003 Mar 29.2990 & 1.0m NOFS    & $V$ & 6$\times$600 & 4.9 & 21.4$\pm$0.2 \\
  &  &  &  & \\
\noalign{\smallskip}
\hline
\noalign{\smallskip}
\end{tabular}
\end{center}
\end{table*}

\newpage

\begin{table*}[th!]
\begin{center}
\begin{tabular}{ccccrc}
\hline
\hline
\noalign{\smallskip}
Mid-exposure & Telescope & Filter & Seeing or &
\multicolumn{1}{c}{Exposure} & Magnitude \\
 time (UT) &  & or grism & slit width ($''$) &
\multicolumn{1}{c}{time (s)} & \\

\noalign{\smallskip}
\hline
\noalign{\smallskip}
\multicolumn{6}{c}{Photometry} \\
\noalign{\smallskip}
\hline
\noalign{\smallskip}

 2003 Mar 28.5296 & 40-inch SSO  & $R$ & 600 & 2.2 & 18.59$\pm$0.04 \\
 2003 Mar 28.5463 & 40-inch SSO  & $R$ & 600 & 2.6 & 18.74$\pm$0.03 \\
 2003 Mar 28.5542 & 40-inch SSO  & $R$ & 600 & 2.4 & 18.81$\pm$0.03 \\
 2003 Mar 28.6305 & 40-inch SSO  & $R$ & 600 & 2.6 & 19.33$\pm$0.05 \\
 2003 Mar 28.6441 & 1.0m ARIES   & $R$ & 200$+$300 & 2.7 & 19.45$\pm$0.11 \\
 2003 Mar 28.6917 & 1.0m ARIES   & $R$ & 300 & 1.9 & 19.53$\pm$0.14 \\
 2003 Mar 28.7570 & 1.0m ARIES   & $R$ & 300 & 2.0 & 19.55$\pm$0.18 \\
 2003 Mar 28.7991 & 1.0m ARIES   & $R$ & 300 & 1.7 & 19.84$\pm$0.14 \\
 2003 Mar 28.9447 & 2.5m NOT & $R$ & 20 &1.0& 20.51$\pm$0.08 \\ 
 2003 Mar 28.9462 & 2.5m NOT & $R$ & 20 &1.0& 20.43$\pm$0.07 \\ 
 2003 Mar 28.9484 & 2.5m NOT & $R$ & 60 &1.0& 20.50$\pm$0.04 \\ 
 2003 Mar 28.9525 & 2.5m NOT & $R$ & 60 &1.0& 20.45$\pm$0.03 \\ 
 2003 Mar 28.9540 & 2.5m NOT & $R$ & 60 &1.0& 20.46$\pm$0.04 \\ 
 2003 Mar 28.9548 & 2.5m NOT & $R$ & 60 &1.0& 20.44$\pm$0.04 \\ 
 2003 Mar 28.9556 & 2.5m NOT & $R$ & 30 &1.0& 20.41$\pm$0.05 \\ 
 2003 Mar 28.9562 & 2.5m NOT & $R$ & 30 &1.0& 20.52$\pm$0.05 \\ 
 2003 Mar 28.9567 & 2.5m NOT & $R$ & 30 &1.0& 20.55$\pm$0.05 \\ 
 2003 Mar 28.9584 & 2.5m NOT & $R$ & 30 &1.0& 20.51$\pm$0.05 \\ 
 2003 Mar 28.9602 & 2.5m NOT & $R$ & 40 &1.0& 20.46$\pm$0.04 \\ 
 2003 Mar 28.9607 & 2.5m NOT & $R$ & 40 &1.0& 20.49$\pm$0.05 \\ 





 2003 Mar 28.9613 & 2.5m NOT & $R$ & 40 &1.0& 20.44$\pm$0.05 \\ 
 2003 Mar 28.9618 & 2.5m NOT & $R$ & 40 &1.0& 20.49$\pm$0.05 \\ 
 2003 Mar 28.9624 & 2.5m NOT & $R$ & 40 &1.0& 20.46$\pm$0.05 \\ 
 2003 Mar 28.9630 & 2.5m NOT & $R$ & 40 &1.0& 20.50$\pm$0.05 \\ 
 2003 Mar 28.9635 & 2.5m NOT & $R$ & 40 &1.0& 20.41$\pm$0.08 \\ 
 2003 Mar 28.9641 & 2.5m NOT & $R$ & 40 &1.0& 20.39$\pm$0.08 \\ 
 2003 Mar 28.9647 & 2.5m NOT & $R$ & 40 &1.0& 20.49$\pm$0.09 \\ 
 2003 Mar 28.9652 & 2.5m NOT & $R$ & 40 &1.0& 20.47$\pm$0.07 \\ 
 2003 Mar 28.9658 & 2.5m NOT & $R$ & 40 &1.0& 20.54$\pm$0.07 \\ 
 2003 Mar 28.9663 & 2.5m NOT & $R$ & 40 &1.0& 20.51$\pm$0.07 \\ 
 2003 Mar 28.9669 & 2.5m NOT & $R$ & 40 &1.0& 20.50$\pm$0.08 \\ 
 2003 Mar 28.9675 & 2.5m NOT & $R$ & 40 &1.0& 20.49$\pm$0.07 \\ 
 2003 Mar 28.9680 & 2.5m NOT & $R$ & 40 &1.0& 20.46$\pm$0.05 \\ 
 2003 Mar 28.9689 & 2.5m NOT & $R$ & 60 &1.0& 20.53$\pm$0.04 \\ 
 2003 Mar 28.9700 & 2.5m NOT & $R$ & 80 &1.0& 20.52$\pm$0.03 \\ 
 2003 Mar 28.9725 & 2.5m NOT & $R$ & 80 &1.0& 20.51$\pm$0.04 \\ 
 2003 Mar 28.9751 & 2.5m NOT & $R$ & 90 &1.0& 20.49$\pm$0.03 \\ 
 2003 Mar 28.9777 & 2.5m NOT & $R$ & 90 &1.0& 20.52$\pm$0.04 \\ 
 2003 Mar 28.9801 & 2.5m NOT & $R$ & 90 &1.0& 20.49$\pm$0.05 \\ 
 2003 Mar 28.9827 & 2.5m NOT & $R$ & 90 &1.0& 20.59$\pm$0.05 \\ 
 2003 Mar 28.9852 & 2.5m NOT & $R$ & 90 &1.0& 20.54$\pm$0.03 \\ 
 2003 Mar 28.9877 & 2.5m NOT & $R$ & 90 &1.0& 20.53$\pm$0.05 \\ 
 2003 Mar 28.9905 & 2.5m NOT & $R$ & 90 &1.0& 20.54$\pm$0.03 \\ 
 2003 Mar 28.9929 & 2.5m NOT & $R$ & 90 &1.0& 20.56$\pm$0.03 \\ 
 2003 Mar 28.9954 & 2.5m NOT & $R$ & 90 &1.0& 20.59$\pm$0.03 \\ 
 2003 Mar 28.9979 & 2.5m NOT & $R$ & 90 &1.0& 20.61$\pm$0.03 \\ 
 2003 Mar 29.0004 & 2.5m NOT & $R$ & 90 &1.0& 20.59$\pm$0.03 \\ 
 2003 Mar 29.0029 & 2.5m NOT & $R$ & 90 &1.0& 20.61$\pm$0.03 \\ 
 2003 Mar 29.0054 & 2.5m NOT & $R$ & 90 &1.0& 20.61$\pm$0.03 \\ 
 2003 Mar 29.0080 & 2.5m NOT & $R$ & 90 &1.0& 20.65$\pm$0.03 \\ 
 2003 Mar 29.0177 & 2.5m NOT & $R$ & 90 &1.0& 20.69$\pm$0.03 \\ 
 2003 Mar 29.0202 & 2.5m NOT & $R$ & 90 &1.0& 20.66$\pm$0.03 \\ 
 2003 Mar 29.0228 & 2.5m NOT & $R$ & 90 &1.0& 20.63$\pm$0.03 \\ 
 2003 Mar 29.0252 & 2.5m NOT & $R$ & 90 &1.0& 20.64$\pm$0.03 \\ 
 2003 Mar 29.0278 & 2.5m NOT & $R$ & 90 &1.0& 20.64$\pm$0.03 \\ 
 2003 Mar 29.0303 & 2.5m NOT & $R$ & 90 &1.0& 20.70$\pm$0.03 \\ 
 2003 Mar 29.0328 & 2.5m NOT & $R$ & 90 &1.0& 20.68$\pm$0.03 \\ 




 2003 Mar 29.0353 & 2.5m NOT & $R$ & 90 &1.0& 20.67$\pm$0.03 \\ 
 2003 Mar 29.0382 & 2.5m NOT & $R$ & 90 &1.0& 20.72$\pm$0.03 \\ 

\noalign{\smallskip}
\hline
\noalign{\smallskip}
\end{tabular}
\end{center}
\end{table*}

\newpage

\begin{table*}[th!]
\begin{center}
\begin{tabular}{ccccrc}
\hline
\hline
\noalign{\smallskip}
Mid-exposure & Telescope & Filter & Seeing or &
\multicolumn{1}{c}{Exposure} & Magnitude \\
 time (UT) &  & or grism & slit width ($''$) &
\multicolumn{1}{c}{time (s)} & \\
\noalign{\smallskip}
\hline
\noalign{\smallskip}
\multicolumn{6}{c}{Photometry} \\
\noalign{\smallskip}
\hline
\noalign{\smallskip}

 2003 Mar 29.1020 & 2.5m NOT & $R$ & 120 &1.0& 20.79$\pm$0.03 \\ 
 2003 Mar 29.1243 & 1.54m Danish & $R$ & 600 & 1.0 & 20.91$\pm$0.09 \\
 2003 Mar 29.1333 & 1.54m Danish & $R$ & 600 & 1.0 & 20.73$\pm$0.07 \\
 2003 Mar 29.1486 & 1.54m Danish & $R$ & 600 & 1.0 & 21.00$\pm$0.09 \\
 2003 Mar 29.1625 & 1.54m Danish & $R$ & 600 & 1.0 & 20.85$\pm$0.06 \\
 2003 Mar 29.1643 & 2.5m NOT & $R$ & 200 &1.0& 20.88$\pm$0.05 \\ 
 2003 Mar 29.1771 & 1.54m Danish & $R$ & 600 & 1.0 & 21.00$\pm$0.07 \\
 2003 Mar 29.1958 & 1.54m Danish & $R$ & 600 & 1.0 & 20.95$\pm$0.07 \\
 2003 Mar 29.2229 & 1.54m Danish & $R$ & 600 & 1.0 & 21.10$\pm$0.07 \\
 2003 Mar 29.2451 & 1.54m Danish & $R$ & 600 & 1.0 & 21.03$\pm$0.07 \\
 2003 Mar 29.2727 & 2.2m ESO/MPG & $R$ & 600 & 0.6 & 21.17$\pm$0.04 \\
 2003 Mar 29.3035 & 1.54m Danish & $R$ & 600 & 1.0 & 21.23$\pm$0.10 \\
 2003 Mar 29.3065 & 1.0m NOFS    & $R$ & 6$\times$600 & 4.8 & 21.2$\pm$0.2 \\
 2003 Mar 29.3236 & 1.54m Danish & $R$ & 600 & 1.0 & 21.36$\pm$0.11 \\
 2003 Mar 31.1056 & 2.5m NOT & $R$ & 2$\times$600 & 1.4 & 22.64$\pm$0.09 \\
  &  &  &  & \\
 2003 Mar 28.6525 & 1.0m ARIES   & $I$ & 2$\times$200 & 2.1 & 18.88$\pm$0.13 \\
 2003 Mar 28.6804 & 1.0m ARIES   & $I$ & 300 & 2.0 & 18.79$\pm$0.29 \\
 2003 Mar 28.7339 & 1.0m ARIES   & $I$ & 300 & 1.9 & 19.25$\pm$0.17 \\
 2003 Mar 28.7459 & 1.0m ARIES   & $I$ & 300 & 1.9 & 19.13$\pm$0.17 \\
 2003 Mar 28.7877 & 1.0m ARIES   & $I$ & 300 & 1.6 & 19.31$\pm$0.22 \\
 2003 Mar 28.8318 & 1.0m ARIES   & $I$ & 300 & 1.6 & 19.41$\pm$0.44 \\
 2003 Mar 29.2438 & 2.2m ESO/MPG & $I$ & 600 & 0.6 & 20.75$\pm$0.05 \\
 2003 Mar 29.3140 & 1.0m NOFS    & $I$ & 6$\times$600 & 4.9 & $>$20.5        \\
  &  &  &  & \\
 2003 Mar 29.0144 & 2.5m NOT     & $i$ & 120 & 1.0 & 20.25$\pm$0.05$^*$ \\

\noalign{\smallskip}
\hline
\noalign{\smallskip}
\multicolumn{6}{c}{Spectroscopy} \\
\noalign{\smallskip}
\hline
\noalign{\smallskip}

 2003 Mar 29.0838 & VLT-$Antu$ & 300V & 6$\times$600 & 1.0 & --- \\

\noalign{\smallskip}
\hline
\noalign{\smallskip}
\multicolumn{6}{c}{Polarimetry} \\
\noalign{\smallskip}
\hline
\noalign{\smallskip}

 2003 Mar 29.2397 & VLT-$Antu$ & $V$ & 5$\times$4$\times$780 & 1.0 & --- \\ 

\noalign{\smallskip}
\hline
\multicolumn{6}{l}{$^*$: calibrated using $I$-band secondary 
standards (see text).} \\
\end{tabular}
\end{center}
\end{table*}


\clearpage

\begin{table}[th!]
\caption[]{Identifications, redshifts, oscillator strengths $f_\lambda$ 
and rest-frame equivalent widths W$_r$ of absorption lines detected at 
the host galaxy redshift in the spectrum of the GRB 030328 afterglow, and 
column densities of the corresponding elements.} 
\begin{center}
\begin{tabular}{lcllc}
\hline
\hline
\noalign{\smallskip}
Line & Redshift & \multicolumn{1}{c}{$f_\lambda$} & 
\multicolumn{1}{c}{W$_r$} & $\log N$ \\
 & & & \multicolumn{1}{c}{(\AA)} & (cm$^{-2}$)  \\
\noalign{\smallskip}
\hline
\noalign{\smallskip}
C {\sc iv}   $\lambda1550$ & 1.5225 & 0.0952 & $0.57\pm0.17$$^{\rm a}$ & --- \\
C {\sc iv}   $\lambda1548$ & 1.5206 & 0.1908 & $1.07\pm0.20$$^{\rm a}$ & --- \\
Al {\sc ii}  $\lambda1670$ & 1.5211 & 1.833  & $0.84\pm0.16$ & $<16.0$ \\
Fe {\sc ii}  $\lambda2600$ & 1.5220 & 0.239  & $0.92\pm0.18$ & $14.3^{+0.6}_{-0.2}$ \\
Fe {\sc ii}  $\lambda2586$ & 1.5209 & 0.0691 & $0.47\pm0.14$ &  $''$ \\
Fe {\sc ii}  $\lambda2382$ & 1.5220 & 0.320  & $0.63\pm0.15$ &  $''$ \\
Fe {\sc ii}  $\lambda2374$ &   ---  & 0.0313 & $<0.4$        &  $''$ \\
Fe {\sc ii}  $\lambda2344$ & 1.5227 & 0.320  & $0.58\pm0.14$ &  $''$ \\
Fe {\sc ii}  $\lambda1608$ &   ---  & 0.058  & $<0.5$        &  $''$ \\
Mg {\sc ii}  $\lambda2803$ & 1.5211 & 0.3054 & $1.34\pm0.19$ & 13.9--15.3 \\
Mg {\sc ii}  $\lambda2796$ & 1.5212 & 0.6123 & $1.57\pm0.20$ &  $''$ \\
Si {\sc ii}  $\lambda1526$ &   ---  & 0.126  & $<0.6$ & $<16.0$ \\
Al {\sc iii} $\lambda1854$ &   ---  & 1.833  & $<0.5$ & $<14.3$ \\
\noalign{\smallskip}
\hline
\noalign{\smallskip}
\multicolumn{4}{l}{$^{\rm a}$The C {\sc iv} doublet is blended.} \\
\end{tabular}
\end{center}
\end{table}

{\bf
\begin{table}[th!]
\caption[]{$P$ and $\theta$ values measured in each of the 5 cycles
of $V$-band polarimetry performed on the GRB030328 afterglow.
$\Delta t$ is the time range of each polarization cycle expressed 
in days after the GRB trigger.}
\begin{center}
\begin{tabular}{cccc}
\hline
\hline
\noalign{\smallskip}
Cycle & $\Delta t$ & $P$ (\%) & $\theta$ ($^{\circ}$)\\ 
\noalign{\smallskip}
\hline
\noalign{\smallskip}
 1 & 0.6570--0.6952 & 2.8$\pm$1.1 & 172$\pm$11 \\
 2 & 0.7165--0.7546 & 2.6$\pm$1.1 & 172$\pm$12 \\
 3 & 0.7564--0.7946 & 1.8$\pm$1.3 & 170$\pm$19 \\
 4 & 0.7981--0.8364 & 2.5$\pm$1.2 & 177$\pm$14 \\
 5 & 0.8386--0.8768 & 2.7$\pm$1.5 & 173$\pm$16 \\
\noalign{\smallskip}
\hline
\noalign{\smallskip}
\end{tabular}
\end{center}
\end{table}
}


\begin{thebibliography}{}


\bibitem{} Assafin, M., Andrei, A.H., Vieira Martins, R., et al. 2001, 
	ApJ, 552, 380

\bibitem{} Barraud, C., Atteia, J.L., Olive, J.-F., Lestrade, et al. 
	2004, Spectral analysis of 50 GRBs detected by {\it HETE-2}, in:
	Gamma-Ray Bursts: 30 Years of Discovery: Gamma-Ray Burst 
	Symposium, ed. E.E. Fenimore, \& M. Galassi (Melville, NY: 
	American Institute of Physics) AIP Conf. Proc., 727, 81

\bibitem{} Bartolini, C., Guarnieri, A., Piccioni, A., Gualandi, R., \&
	Pizzichini, G. 2003, GCN Circ. 2008

\bibitem{} Beuermann, K., Hessman, F.V., Reinsch, K., et al. 1999, A\&A, 
	352, L26

\bibitem{} Bj\"ornsson, G., Gudmundsson, E.H., \& J\'ohannesson, G. 2004,
	ApJ, 615, L77

\bibitem{} Bloom, J.S., Frail, D.A., \& Sari, R. 2001, AJ, 121, 2879

\bibitem{} Burenin, R., Denissenko, D., Pavlinsky, M., et al. 2003,
	GCN Circ. 1990



\bibitem{} Butler, N.R., Ricker, G. R., Ford, P.G., et al. 2005, ApJ, 
	629, 908           

\bibitem{} Cardelli, J.A., Clayton, G.C., \& Mathis, J.S. 1989, ApJ, 345, 
	245

\bibitem{} Chevalier, R.A., \& Li, Z.Y. 2000, ApJ, 536, 195

\bibitem{} Costa, E., Frontera, F., Heise, J., et al. 1997, Nature, 387, 
	783

\bibitem{} Covino, S., Lazzati, D., Ghisellini, G., et al. 1999, 348, L1

\bibitem{} Dai, Z.G., \& Cheng, K.S. 2001, ApJ, 558, L109 

\bibitem{} de Ugarte Postigo, A., Castro-Tirado, A.J., Gorosabel, J.,
	et al. 2005, A\&A, 443, 841

\bibitem{} Deutsch, E.W. 1999, AJ, 118, 1882

\bibitem{} Di Serego Alighieri, S. 1997, Polarimetry with large 
	telescopes, in: Instrumentation for large telescopes, ed. J.M. 
	Rodriguez Espinosa, A. Herrero, \& F. S\'anchez (Cambridge: 
	Cambridge Univ. Press), p. 287

\bibitem{} Dickey, J.M., \& Lockman, F.J. 1990, ARA\&A, 28, 215

\bibitem{} Fiore, F., D'Elia, V., Lazzati, D., et al. 2005, ApJ, 624, 853 

\bibitem{} Frail, D.A., Kulkarni, S.R., Nicastro, L., Feroci, M., \& 
	Taylor, G.B. 1997, Nature, 389, 261


\bibitem{} Fugazza, D., Fiore, F., Cocchia, M., et al. 2003a, GCN Circ.
        1983

\bibitem{} Fugazza, D., Antonelli, L.A., Fiore, F., et al. 2003b, GCN 
	Circ. 1982

\bibitem{} Fukugita, M., Shimasaku, K., \& Ichikawa, T. 1995, PASP, 107,
        945

\bibitem{} Galama, T.J., Wijers, R.A.M.J., Bremer, M., et al. 1998a, ApJ, 
	500, L97

\bibitem{} Galama, T.J., Vreeswijk, P.M., van Paradijs, J., et al. 1998b,
	Nature, 395, 670

\bibitem{} Gal-Yam, A., Ofek, E.O., \& Polishook, D. 2003, GCN Circ. 1984

\bibitem{} Garnavich, P., Martini, P., \& Stanek, K.Z. 2003, GCN Circ.
	2036

\bibitem{} Ghirlanda, G., Ghisellini, G., \& Lazzati, D. 2004, ApJ, 616, 
	331

\bibitem{} Ghisellini, G., \& Lazzati, D. 1999, MNRAS, 309, L7

\bibitem{} Gorosabel, J., Jel\'{i}nek, M., de Ugarte Postigo, A., Guziy, S.,
	Castro-Tirado, A.J. et al. 2005, The GRB 030328 host: Another case 
	of a blue starburst galaxy, in: 4th Workshop - Gamma-Ray Burst 
	in the Afterglow Era, ed. L. Piro, L. Amati, S. Covino, \& B. 
	Gendre, Il Nuovo Cimento, 28C, 677

\bibitem{} Greiner, J., Klose, S., Reinsch, K., et al. 2003a, Nature, 426, 
	157

\bibitem{} Greiner, J., Peimbert, M., Estaban, C., et al. 2003b, GCN Circ. 
	2020

\bibitem{} Guarnieri, A., Bartolini, C., Masetti, N., et al. 1997, A\&A,
	328, L13 

\bibitem{} Guidorzi, C., Monfardini, A., Gomboc, A., et al. 2005, ApJ,
	630, L121

\bibitem{} Hamuy, M., Walker, A.R., Suntzeff, N.B. et al. 1992, PASP, 104, 
	533 

\bibitem{} Hamuy, M., Suntzeff, N.B., Heathcote, S.R. et al. 1994, PASP, 
	106, 566 

\bibitem{} Henden, A.A. 2003, GCN Circ. 2114

\bibitem{} Hjorth, J., Sollerman, J., M\o ller, P., et al. 2003, Nature, 
	423, 847

\bibitem{} Ibrahimov, M.A., Asfandiyarov, I.M., Kahharov, B.B., et al. 2003,
	GCN Circ. 2192

\bibitem{} Iwamoto, K., Mazzali, P.A., Nomoto, K., et al. 1998, Nature, 395, 
	672

\bibitem{} Kouveliotou, C., Meegan, C.A., Fishman, G.J., et al. 1993, ApJ, 
	413, L101

\bibitem{} Kumar, P., \& Piran, T. 2000, ApJ, 532, 286

\bibitem{} Lazzati, D., Covino, S., Di Serego Alighieri, S., et al. 2003,
	A\&A, 410, 823

\bibitem{} Malesani, D., Tagliaferri, G., Chincarini, G., et al. 2004, ApJ,
        609, L5

\bibitem{} Martini, P., Garnavich, P., \& Stanek, K.Z. 2003a, GCN Circ. 1980

\bibitem{} Martini, P., Garnavich, P., \& Stanek, K.Z. 2003b, GCN Circ. 1979

\bibitem{} Masetti, N., Pian, E., Palazzi, E., et al. 1999, A\&AS, 138, 453

\bibitem{} Masetti, N., Bartolini, C., Bernabei, S., et al. 2000, A\&A, 359, 
	L23 

\bibitem{} Matheson, T., Garnavich, P.M., Stanek, K.Z., et al. 2003, ApJ,
	599, 394

\bibitem{} M\'esz\'aros, P., \& Rees, M.J. 1999, MNRAS, 306, L39

\bibitem{} Mirabal, N., Halpern, J.P., Chornock, R., et al. 2003, ApJ, 
	595, 935

\bibitem{} Nestor, D.B., Turnshek, D.A., Rao, S.M., et al. 2005, ApJ, 628, 
	637

\bibitem{} Palazzi, E., Pian, E., Masetti, N., et al. 1998, A\&A, 336, L95

\bibitem{} Peterson, B.A., \& Price, P.A. 2003, GCN Circ. 1974


\bibitem{} Price, P.A., \& Peterson, B.A. 2003, GCN Circ. 1977

\bibitem{} Rao, S.M., \& Turnshek, D.A. 2000, ApJS, 130, 1

\bibitem{} Ramaprakash, A.N. 1998, Ph.D. Thesis, Inter-University Centre
        for Astronomy and Astrophysics
        ({\tt http://www.iucaa.ernet.in/\~{}anr/thesis.html})

\bibitem{} Rhoads, J.E. 1999, ApJ, 525, 737

\bibitem{} Rol, E., Wijers, R.A.M.J., Vreeswijk, P.M., et al. 2000, ApJ,
	544, 707

\bibitem{} Rol, E., Wijers, R.A.M.J., Fynbo, J.P.U., et al. 2003a, A\&A, 
	405, L23

\bibitem{} Rol, E., Vreeswijk, P., \& Jaunsen, A. 2003b, GCN Circ. 1981

\bibitem{} Rossi, E.M., Lazzati, D., Salmonson, J.D., Ghisellini, G. 
	2004, MNRAS, 354, 86

\bibitem{} Rumyantsev, V., Biryukov, V., \& Pozanenko, A. 2003, GCN 
	Circ. 1991

\bibitem{} Sari, R. 1999, ApJ, 524, L43

\bibitem{} Sari, R., Piran, T., \& Narayan, R. 1998, ApJ, 497, L17

\bibitem{} Sari, R., Piran, T., \& Halpern, J.P. 1999, ApJ, 519, L17

\bibitem{} Savaglio, S., \& Fall, S.M. 2004, ApJ, 614, 293

\bibitem{} Savaglio, S., \& Fall, S.M., \& Fiore, F. 2003, ApJ, 585, 638

\bibitem{} Schlegel, D.J., Finkbeiner, D.P. \& Davis, M. 1998, ApJ, 500, 525

\bibitem{} Simmons, J.F.L., \& Stewart B.G. 1985, A\&A, 142, 100

\bibitem{} \v{S}imon, V., Hudec, R., Pizzichini, G., \& Masetti, N. 
	2001, A\&A, 377, 450

\bibitem{} Stanek, K.Z., Matheson, T., Garnavich, P.M., et al. 2003, 
	ApJ, 591, L17 

\bibitem{} Stetson, P.B. 1987, PASP, 99, 191

\bibitem{} Vanderspek, R., Crew, G., Doty J., et al. 2003, GCN Circ. 1997

\bibitem{} van Dokkum, P.G. 2001, PASP, 113, 1420

\bibitem{} van Paradijs, J., Groot, P.J., Galama, T.J., et al. 1997,
        Nature, 386, 686

\bibitem{} Villasenor, J., Crew, G., Vanderspek, R., et al. 2003, GCN circ. 
	1978

\bibitem{} Vreeswijk, P.M., Fruchter, A., Kaper, L., et al. 2001, ApJ, 546, 
	672

\bibitem{} Vreeswijk, P.M., Ellison, S., Ledoux, C., et al. 2004, A\&A, 419, 
	927

\bibitem{} Wardle, J.F.C., \& Kronberg, P.P. 1974, ApJ, 194, 249

\bibitem{} Wijers, R.A.M.J., Vreeswijk, P.M., Galama, T.J., et al. 1999,
        ApJ, 523, L33

\bibitem{} Woosley S.E., \& Heger, A. 2006, ApJ, 637, 914

\end{thebibliography}
\end{document}